\documentclass[12pt]{article}
  
\usepackage{amsmath,amssymb}
\usepackage{bm}
\usepackage{graphicx, color}
\usepackage{wrapfig}
\usepackage{cite}
\begin{document}
\title{
\begin{flushright}
\ \\*[-80pt] 
\begin{minipage}{0.2\linewidth}
\normalsize
EPHOU-18-009 \\
HUPD1807 \\*[50pt]
\end{minipage}
\end{flushright}
{\Large \bf 
 Modular $A_4$ invariance and neutrino mixing 
\\*[20pt]}}

\author{ 
\centerline{
~Tatsuo Kobayashi $^{1}$,
~~Naoya Omoto $^{1}$,
~~Yusuke~Shimizu $^{2}$,} \\*[5pt]
\centerline{~Kenta~Takagi $^{2}$,~~Morimitsu Tanimoto $^{3}$, 
~~Takuya H. Tatsuishi $^{1}$} 
\\*[20pt]
\centerline{
\begin{minipage}{\linewidth}
\begin{center}
$^1${\it \normalsize
Department of Physics, Hokkaido University, Sapporo 060-0810, Japan} \\*[5pt]
$^2${\it \normalsize
Graduate School of Science, Hiroshima University, Higashi-Hiroshima 739-8526, Japan} \\*[5pt]
$^3${\it \normalsize
Department of Physics, Niigata University, Niigata 950-2181, Japan}
\end{center}
\end{minipage}}
\\*[50pt]}

\date{
\centerline{\small \bf Abstract}
\begin{minipage}{0.9\linewidth}
\medskip 
\medskip 
\small 
We study the phenomenological implications of the modular symmetry $\Gamma(3) \simeq A_4$ of lepton flavors facing recent experimental data of neutrino oscillations.
The mass matrices of neutrinos and charged leptons are essentially given by fixing the expectation value of modulus $\tau$, which is the only source of modular invariance breaking.
We introduce no flavons in contrast with the conventional flavor models with $A_4$ symmetry.
We classify our neutrino models along with the type I seesaw model, the Weinberg operator model and the Dirac neutrino model.
In the normal hierarchy of neutrino masses, the seesaw model is available by taking account of recent experimental data of neutrino oscillations and the cosmological bound of sum of neutrino masses.
The predicted $\sin^2\theta_{23}$ is restricted to be larger than $0.54$ and $\delta_{CP}=\pm (50^{\circ}\mbox{--}180^{\circ})$.
Since the correlation of $\sin^2\theta_{23}$ and $\delta_{CP}$ is sharp, the prediction is testable in the future.
It is remarkable that the effective mass $m_{ee}$ of the neutrinoless double beta decay is around $22$\,meV while the sum of neutrino masses is predicted to be $145$\,meV.
On the other hand, for the inverted hierarchy of neutrino masses, 
only the Dirac neutrino model is  consistent with the experimental data.
\end{minipage}
}

\begin{titlepage}
\maketitle
\thispagestyle{empty}
\end{titlepage}

\section{Introduction}

In spite of the remarkable success of the standard model (SM),
the origin of the flavors of quarks and leptons is still unknown.
The recent developments of the neutrino oscillation experiments provide us important clues to investigate the flavor physics.
Indeed, the neutrino oscillation experiments have determined two neutrino mass squared differences and three neutrino  mixing angles precisely.
In particular, the recent data of both T2K~\cite{Abe:2017vif, T2K} and NO$\nu$A~\cite{Adamson:2017gxd, nova} imply that the atmospheric neutrino mixing angle $\theta_{23}$ is expected to be near the maximal angle $45^\circ$.
The closer the observed $\theta_{23}$ is to the maximal mixing, the more likely some flavor symmetry is to exist behind it.
In addition to the precise measurements of neutrino mixing angles,
T2K and NO$\nu$A strongly indicate CP violation in the neutrino oscillation~\cite{T2K,nova}.
Thus, we are in the era to develop the flavor theory of leptons with facing both the flavor mixing angles and CP violating phase.

One of interesting approaches is to impose non-Abelian discrete symmetries for flavors.
Many models have been proposed by using $S_3$, $A_4$, $S_4$, $A_5$ and other groups with lager orders \cite{Altarelli:2010gt,Ishimori:2010au,Ishimori:2012zz,King:2013eh,King:2014nza}.
In particular, $A_4$ flavor models are attractive because
$A_4$ is the minimal group which has a triplet as its irreducible representation
and enable us to explain three families of quarks and leptons naturally
\cite{Ma:2001dn,Babu:2002dz,Altarelli:2005yp,Altarelli:2005yx,Shimizu:2011xg,Kang:2018txu}.
However, variety of models is so wide that it is difficult to obtain clear clues of flavor symmetry.
Indeed, symmetry breakings are required to reproduce realistic mixing angles
\cite{Petcov:2018snn}.
The effective Lagrangian of a typical flavor model is given by introducing the gauge singlet scalars which are so-called flavons.
Those vacuum expectation values (VEVs) determine the flavor structure of quarks and leptons.
Finally, the breaking sector of flavor symmetry typically produces many unknown parameters.

Superstring theory with certain compactifications can lead to non-Abelian discrete flavor symmetries.
For example, heterotic orbifold models lead to $D_4$, $\Delta(54)$, etc. \cite{Kobayashi:2006wq}.
(See also \cite{Kobayashi:2004ya,Ko:2007dz}.)
Similar flavor symmetries are also derived in type II magnetized and intersecting D-brane models 
\cite{Abe:2009vi,BerasaluceGonzalez:2012vb}.
On the other hand, string theory on tori or orbifolds has the modular symmetry which 
acts non-trivially on flavors of quarks and leptons
\cite{Lauer:1989ax,Lerche:1989cs,Ferrara:1989qb,Cremades:2004wa,Kobayashi:2017dyu,Kobayashi:2018rad}.
In this sense, the modular symmetry is a non-Abelian discrete flavor symmetry.

It is interesting that the modular group includes $S_3$, $A_4$, $S_4$, and $A_5$
as its finite subgroups, $\Gamma(N)$.
However, there is a difference between the modular symmetry and the usual flavor symmetry.
Yukawa couplings are written as modular forms, functions of the modulus $\tau$,
and transform non-trivially under the modular symmetry as well as fields.
On the other hand, Yukawa couplings are invariants in the usual flavor symmetries.
In this aspect, an attractive ansatz was proposed by taking  $\Gamma(3) \simeq A_4$ 
in Ref.\cite{Feruglio:2017spp}
where Yukawa couplings are $A_4$ triplets of modular forms, and 
both left-handed leptons and right-handed neutrinos are $A_4$ triplets while right-handed charged leptons are $A_4$ singlets.
Along with this work, $\Gamma(2) \simeq S_3$ \cite{Kobayashi:2018vbk}
and $\Gamma(4) \simeq S_4$ \cite{Penedo:2018nmg} have been discussed
as well as the numerical works \cite{Criado:2018thu}.
These are bottom-up approaches for model building, but not a top-down approach from explicit string models.
However, these approaches would make a bridge between neutrino physics and underlying theory such as
superstring theory from the viewpoint of flavor symmetries.

In this paper, we present a comprehensive study of $\Gamma(3) \simeq A_4$ numerically
by taking account of the recent experimental data of neutrino oscillations.
The mass matrices of neutrinos and charged leptons
are essentially given by the expectation value of the modulus $\tau$,
which is the only source of modular invariance breaking.
However, there are freedoms for the assignments of irreducible representations and modular weights to leptons.
We study neutrino mass matrices for three classified models:
the type I seesaw model, the Weinberg operator model, and the Dirac neutrino model.
In order to build models with minimal number of parameters, we introduce no flavons.


The paper is organized as follows.
In section 2, we give a brief review on modular symmetry. 
In section 3, we present the mass matrices for neutrinos and charged leptons in our models.
In section 4, we present the numerical results of our models.
Section 5 is devoted to a summary.
Appendix A shows the relevant multiplication rules of $A_4$. 
In Appendix B, we show how to determine the coupling coefficients of the charged lepton sector.
Appendix C shows the lepton mixing matrix and the relevant measures which are used in this work for the case of Majorana neutrinos.

\vskip 1 cm
\section{Modular symmetry}

In this section, we give a brief review on the modular symmetry on the torus and 
its low-energy effective field theory.

The torus compactification is the simplest compactification.
For example, the two-dimensional torus $T^2$ can be constructed as
division of $\mathbb{R}^2$ by a two-dimensional lattice $\Lambda$, i.e. 
$T^2=\mathbb{R}^2/\Lambda$.
Here, we use the complex coordinate on $\mathbb{R}^2$ with the lattice spanned by two lattice vectors,
$\alpha_1=2 \pi R$ and $\alpha_2 = 2 \pi R \tau$;
where $R$ is real and $\tau$ is a complex modulus parameter.
However, there is some ambiguity in choice of the basis vectors.
The same lattice can be spanned by the following basis vectors,
\begin{equation}
\label{eq:SL2Z}
    \left(
    \begin{array}{c}
    \alpha'_2 \\ \alpha'_1
    \end{array}
    \right) =\left(
    \begin{array}{cc}
    a & b \\
    c & d   
    \end{array}
    \right) \left(
    \begin{array}{c}
    \alpha_2 \\ \alpha_1
    \end{array}
    \right) \ ,
\end{equation}
where $a,b,c,d$ are integer with satisfying $ad-bc = 1$.
That is the $SL(2,\mathbb{Z})$ transformation.
Under the above transformation, the modulus parameter transforms as 
\begin{equation}\label{eq:tau-SL2Z}
    \tau \longrightarrow \tau'= \frac{a\tau + b}{c \tau + d}\ ,
\end{equation}
and this modular transformation is generated by $S$ and $T$, 
\begin{eqnarray}
    & &S:\tau \longrightarrow -\frac{1}{\tau}\ , \\
    & &T:\tau \longrightarrow \tau + 1\ .
\end{eqnarray}
They satisfy the following algebraic relations, 
\begin{equation}
    S^2 =\mathbb{I}\ , \qquad (ST)^3 =\mathbb{I}\ .
\end{equation}
If we impose $T^N=\mathbb{I}$ furthermore, 
we obtain finite subgroups $\Gamma(N)$.
$\Gamma(N)$ with $N=2,3,4,5$ are isomorphic to
$S_3$, $A_4$, $S_4$ and $A_5$, respectively \cite{deAdelhartToorop:2011re}.
Indeed, $\Gamma(N)$ is a quotient of the modular group by the so-called congruence subgroup $\bar{\Gamma}(N)$.
Holomorphic functions which transform as
\begin{equation}
    f(\tau)\to (c\tau+d)^kf(\tau)~,
\end{equation}
under the modular transformation Eq.(\ref{eq:tau-SL2Z}) are called modular forms of weight $k$.

Superstring theory on the torus $T^2$ or orbifold $T^2/Z_N$ has the modular symmetry.
Its low-energy effective field theory is described in terms of  supergravity theory,
and  string-derived supergravity theory has also the modular symmetry.
Under the modular transformation Eq.(\ref{eq:tau-SL2Z}), chiral superfields $\phi^{(I)}$ 
transform as \cite{Ferrara:1989bc},
\begin{equation}
    \phi^{(I)}\to(c\tau+d)^{-k_I}\rho^{(I)}(\gamma)\phi^{(I)},
\end{equation}
where  $-k_I$ is the so-called modular weight and $\rho^{(I)}(\gamma)$ denotes a unitary representation matrix of $\gamma\in\Gamma(N)$.
The kinetic terms of their scalar components are written by 
\begin{equation}
    \sum_I\frac{|\partial_\mu\phi^{(I)}|^2}{\langle-i\tau+i\bar{\tau}\rangle^{k_I}} ~,
    \label{kinetic}
\end{equation}
which is invariant under the modular transformation.
Here, we use the convention that the superfield and its scalar component are denoted by the same letter.
Also, the superpotential should be invariant under the modular symmetry.
That is, the superpotential should have vanishing modular weight in global supersymmetric models, 
while the superpotential in supergravity should be invariant under the modular symmetry up to the K\"ahler 
transformation.
In the following sections, we study global supersymmetric models, e.g. minimal supersymmetric standard model 
(MSSM) and its extension with right-handed neutrinos.
Thus, the superpotential has vanishing modular weight.
However, note that Yukawa couplings as well as higher order couplings depend on modulus, 
and they can have non-vanishing modular weights.
The breaking scale of supersymmetry can be between $\mathcal{O}(1)$TeV and the compactification scale.
The modular symmetry is broken by the vacuum expectation value of $\tau$, i.e. at the compactification scale, 
which is of order of the Planck scale or slightly lower scale.

The Dedekind eta-function $\eta(\tau)$ is one of famous modular forms, which is written by 
\begin{equation}
\eta(\tau) = q^{1/24} \prod_{n =1}^\infty (1-q^n)~,
\end{equation}
where $q = e^{2 \pi i \tau}$ and $\eta(\tau)^{24}$ is a modular form of weight~12.
By use of  $\eta(\tau)$ and its derivative,  $A_4$ triplet modular forms $(Y_1,Y_2,Y_3)$ 
of modular weight~2 are written by \cite{Feruglio:2017spp},
\begin{eqnarray} 
\label{eq:Y-A4}
Y_1(\tau) &=& \frac{i}{2\pi}\left( \frac{\eta'(\tau/3)}{\eta(\tau/3)}  +\frac{\eta'((\tau +1)/3)}{\eta((\tau+1)/3)}  
+\frac{\eta'((\tau +2)/3)}{\eta((\tau+2)/3)} - \frac{27\eta'(3\tau)}{\eta(3\tau)}  \right), \nonumber \\
Y_2(\tau) &=& \frac{-i}{\pi}\left( \frac{\eta'(\tau/3)}{\eta(\tau/3)}  +\omega^2\frac{\eta'((\tau +1)/3)}{\eta((\tau+1)/3)}  
+\omega \frac{\eta'((\tau +2)/3)}{\eta((\tau+2)/3)}  \right) , \label{tripletY} \\ 
Y_3(\tau) &=& \frac{-i}{\pi}\left( \frac{\eta'(\tau/3)}{\eta(\tau/3)}  +\omega\frac{\eta'((\tau +1)/3)}{\eta((\tau+1)/3)}  
+\omega^2 \frac{\eta'((\tau +2)/3)}{\eta((\tau+2)/3)}  \right) , \nonumber
\end{eqnarray}
where 
$\omega= e^{2\pi i/3}$.
The overall coefficient in Eq.(\ref{tripletY}) is one choice and cannot be determined essentially.

\section{Models with modular symmetry}
Let us consider a modular invariant flavor model with the $A_4$ symmetry for leptons.
At first, we discuss the type I seesaw model where neutrinos are Majorana particles.
There are freedoms for the assignments of irreducible representations and modular weights to leptons.
We suppose that 
three left-handed lepton doublets are compiled in a triplet of $A_4$.
The three right-handed neutrinos are also of a triplet of $A_4$.
On the other hand, the Higgs doublets are supposed to be singlets of $A_4$.
The generic assignments of representations and modular weights to the MSSM fields and right-handed neutrino superfields 
are presented in Table \ref{tb:fields}. 
In order to build a model with minimal number of parameters, we introduce no flavons.

For the charged leptons, we assign three right-handed charged leptons for three different singlets of $A_4$, $(1,1'',1')$.
Therefore, there are three independent couplings in the superpotential of the charged lepton sector.
Those coupling constants can be adjusted to the observed charged lepton masses.
Since there are three singlets in the $A_4$ group,
there are six cases for the assignment of three right-handed charged leptons.
However, the freedom of these assignments for right-handed charged leptons do not affect the results for lepton mixing angles.
    
It may be helpful to comment that if the right-handed charged leptons are of a $A_4$ triplet, 
we cannot reproduce the well known charged lepton mass hierarchy $1:\lambda^2 : \lambda^5$,
where $\lambda\simeq 0.2$.
\begin{table}[h]
	\centering
	\begin{tabular}{|c||c|c|c|c|c|c|} \hline 
		&$L$&$e_R,\mu_R,\tau_R$&$\nu_R$&$H_u$&$H_d$&$Y$\\ \hline \hline 
		\rule[14pt]{0pt}{0pt}
		$SU(2)$&$2$&$1$&$1$&$2$&$2$&$1$\\
		$A_4$&$3$& $1$,\ $1''$,\ $1'$&$3$&$1$&$1$&$3$\\
		$-k_I$&$-1\ (1)$&$-1\ (-3)$&$-1$&0&0&$k=2$ \\ \hline
	\end{tabular}
	\caption{
		The charge assignment of $SU(2)$, $A_4$, and the modular weight ($-k_I$ for fields and $k$ for coupling $Y$)
		 in the type I seesaw model.
		The right-handed charged leptons are assigned three $A_4$ singlets, respectively.  Values of $-k_I$ in the parentheses
		are alternative assignments of the modular weight.}
	\label{tb:fields}
\end{table}

The modular invariant mass terms of the leptons are given as the following superpotentials:
\begin{align}
w_e&=\alpha e_RH_d(LY)+\beta \mu_RH_d(LY)+\gamma \tau_RH_d(LY)~,\label{charged} \\
w_D&=g(\nu_R H_u L Y)_{\bf 1}~,  \label{Dirac}\\
w_N&=\Lambda(\nu_R\nu_RY)_{\bf 1}~, \label{Majorana}
\end{align}
where sums of the modular weights  vanish.
The parameters $\alpha$,  $\beta$,  $\gamma$,  $g$, and $\Lambda$
 are constant coefficients.
The functions $Y_i(\tau)$ are   $A_4$ triplet modular forms
 and they consist of the modulus parameter~$\tau$:
\begin{align}
Y=\begin{pmatrix}Y_1(\tau)\\Y_2(\tau)\\Y_3(\tau)\end{pmatrix}=
\begin{pmatrix}
1+12q+36q^2+12q^3+\dots \\
-6q^{1/3}(1+7q+8q^2+\dots) \\
-18q^{2/3}(1+2q+5q^2+\dots)\end{pmatrix},\qquad q=e^{2\pi i \tau},
\end{align}
where the $q$-expansion of $Y_i(\tau)$ is used.
The $Y_i(\tau)$ satisfy the constraint \cite{Feruglio:2017spp}:
\begin{align}
Y_2^2+2Y_1Y_3=0~.
\end{align}
Since the dimension of the space of modular forms of weight~2 for $\Gamma(3)\simeq A_4$ is 3 (see, e.g.  \cite{Feruglio:2017spp,Gunning:1962}), all $Y$'s in Eqs.(\ref{charged})-(\ref{Majorana}) are the same modular forms.

There is an alternative assignment of the modular weight for the left-handed lepton and the right-handed charged leptons as presented in parentheses of Table \ref{tb:fields}
\cite{Criado:2018thu}.
For the alternative assignment, the modular invariant superpotential $w_D$ is given with constant parameters
without the modular coupling $Y$ as:
\begin{align}
w_D&=g(\nu_R H_u L)_{\bf 1}~.  \label{Dirac2}
\end{align}

Next, we discuss the case where neutrino masses originate from the Weinberg operator.
We have the unique possibility of the superpotential
\begin{align}
w_\nu&=-\frac{1}{\Lambda}(H_u H_u LLY)_{\bf 1}~,
\label{Weinberg}
\end{align}
where both modular weights of  $L$ and  right-handed charged leptons are $-1$ as shown  in Table \ref{tb:fields}.

There is another possibility for neutrinos, that is, neutrinos are Dirac particles.
In this case, the neutrino mass matrix is derived only from $w_D$ in Eq.(\ref{Dirac}).

\subsection{Charged lepton mass matrix}


Let us consider an assignment of $A_4$ for the right-handed charged leptons 
as $(e_R,\mu_R,\tau_R)=(1,1'',1')$ in Table~\ref{tb:fields}.
By using the decomposition rule of a $A_4$ tensor product in Appendix A, 
we obtain the mass matrix of charged leptons as follows
\footnote{There are six cases to assign $A_4$ singlets for the right-handed charged leptons as 
$(e_R,\mu_R,\tau_R)=(1,1'',1')$, $(1,1',1'')$, $(1',1,1'')$, $(1',1'',1)$, $(1'',1',1)$, $(1'',1,1')$. 
The mass matrices are obtained by permutations of rows each other.
 Then, the combinations $M_E^\dagger M_E$ are  same ones  up to re-labeling of
  parameters $\alpha$, $\beta$, and $\gamma$ for all cases.}:

\begin{align}
\begin{aligned}
M_E&={\rm diag}[\alpha, \beta, \gamma]
\begin{pmatrix}
Y_1 & Y_3 & Y_2 \\
Y_2 & Y_1 & Y_3 \\
Y_3 & Y_2 & Y_1
\end{pmatrix}_{RL}.
\end{aligned}\label{eq:CL}
\end{align}



The coefficients $\alpha$, $\beta$, and $\gamma$ 
are taken to be real positive by rephasing  right-handed charged lepton fields
without loss of generality.
Those parameters can be written in terms of the modulus parameter $\tau$ and the charged lepton masses as seen in Appendix B.

\subsection{Neutrino mass matrix}
Since the tensor product of $3\otimes 3$ is decomposed into a symmetric triplet and an antisymmetric triplet as seen in Appendix A,  the superpotential of
the Dirac neutrino mass  in Eq.(\ref{Dirac}) is expressed with additional two parameters $g_1$ and $g_2$ as:
\begin{align}
\begin{aligned}
w_D=&v_u\begin{pmatrix}\nu_{R1}\\\nu_{R2}\\\nu_{R3}\end{pmatrix} \otimes\left[
g_1\begin{pmatrix}
2\nu_eY_1-\nu_\mu Y_3-\nu_\tau Y_2\\
2\nu_\tau Y_3-\nu_e Y_2-\mu Y_1\\
2\nu_\mu Y_2-\nu_\tau Y_1-\nu_eY_3\end{pmatrix} \oplus
g_2\begin{pmatrix}\nu_\mu Y_3-\nu_\tau Y_2\\\nu_e Y_2-\nu_\mu Y_1\\\nu_\tau Y_1-\nu_e Y_3\end{pmatrix}\right] \\
=&v_ug_1\left[
\nu_{R1}(2\nu_eY_1-\nu_\mu Y_3-\nu_\tau Y_2)+
\nu_{R2}(2\nu_\mu Y_2-\nu_\tau Y_1-\nu_eY_3)+
\nu_{R3}(2\nu_\tau Y_3-\nu_e Y_2-\nu_\mu Y_1)\right] \\
&+v_ug_2\left[
\nu_{R1}(\nu_\mu Y_3-\nu_\tau Y_2)+
\nu_{R2}(\nu_\tau Y_1-\nu_e Y_3)+
\nu_{R3}(\nu_e Y_2-\nu_\mu Y_1)\right].
\end{aligned}
\end{align}
The  Dirac neutrino mass matrix is given as
\begin{align}
M_D=v_u\begin{pmatrix}
2g_1Y_1 & (-g_1+g_2)Y_3 & (-g_1-g_2)Y_2 \\
(-g_1-g_2)Y_3 & 2g_1Y_2 & (-g_1+g_2)Y_1 \\
(-g_1+g_2)Y_2 & (-g_1-g_2)Y_1 & 2g_1Y_3\end{pmatrix}_{RL}.
\label{MD}
\end{align}
 
 For the alternative case in Eq.(\ref{Dirac2}), the superpotential
 of the Dirac neutrino  is written as:
 \begin{align}
 \begin{aligned}
 w_D= v_u g\begin{pmatrix}\nu_{R1}\\\nu_{R2}\\\nu_{R3}\end{pmatrix} \otimes
 \begin{pmatrix}\nu_{e}\\\nu_{\mu}\\\nu_{\tau}\end{pmatrix} 
 =v_u g\left (\nu_{R1}\nu_e+ \nu_{R2}\nu_\tau +\nu_{R3}\nu_\mu \right ).
 \end{aligned}
 \end{align}
 The Dirac neutrino mass matrix is simply  given as
 \begin{align}
 M_D=v_u g\begin{pmatrix}
 1 & 0 &0 \\  0 & 0 & 1 \\ 0& 1 & 0\end{pmatrix}_{RL}.
 \label{Dirac-constant}
 \end{align}

On the other hand, since the Majorana neutrino mass terms are symmetric,
the superpotential in Eq.(\ref{Majorana}) is expressed simply as
\begin{align}
\begin{aligned}
w_N=&\Lambda\begin{pmatrix}
2\nu_{R1}\nu_{R1}-\nu_{R2}\nu_{R3}-\nu_{R3}\nu_{R2}\\
2\nu_{R3}\nu_{R3}-\nu_{R1}\nu_{R2}-\nu_{R2}\nu_{R1}\\
2\nu_{R2}\nu_{R2}-\nu_{R3}\nu_{R1}-\nu_{R1}\nu_{R3}\end{pmatrix}\otimes
\begin{pmatrix}Y_{1}\\Y_{2}\\Y_{3}\end{pmatrix} \\
=&\Lambda\left[(2\nu_{R1}\nu_{R1}-\nu_{R2}\nu_{R3}-\nu_{R3}\nu_{R2})Y_1+
(2\nu_{R3}\nu_{R3}-\nu_{R1}\nu_{R2}-\nu_{R2}\nu_{R1})Y_3\right. \\
&\left.+(2\nu_{R2}\nu_{R2}-\nu_{R3}\nu_{R1}-\nu_{R1}\nu_{R3})Y_2\right].
\end{aligned}
\end{align}
Then, the right-handed Majorana neutrino mass matrix is given as
\begin{align}
M_N=\Lambda\begin{pmatrix}
2Y_1 & -Y_3 & -Y_2 \\
-Y_3 & 2Y_2 & -Y_1 \\
-Y_2 & -Y_1 & 2Y_3\end{pmatrix}_{RR}.
\label{MajoranaR}
\end{align}
Finally, the effective neutrino mass matrix is obtained by the type I seesaw 
as follows:
\begin{align}
M_\nu=-M_D^{\rm T}M_N^{-1}M_D ~.
\end{align}

\begin{table}[h]
	\centering
	\begin{tabular}{|c||c|} \hline 
		Models	&Mass Matrices\\ \hline
		\hskip -1.7 cm I (a) :  Seesaw&
		{\scriptsize $ M_D\sim\begin{pmatrix}
			2g_1Y_1 & (-g_1+g_2)Y_3 & (-g_1-g_2)Y_2 \\
			(-g_1-g_2)Y_3 & 2g_1Y_2 & (-g_1+g_2)Y_1 \\
			(-g_1+g_2)Y_2 & (-g_1-g_2)Y_1 & 2g_1Y_3\end{pmatrix} $}, \quad
		{\scriptsize	$M_N\sim\begin{pmatrix}
			2Y_1 & -Y_3 & -Y_2 \\
			-Y_3 & 2Y_2 & -Y_1 \\
			-Y_2 & -Y_1 & 2Y_3\end{pmatrix}$}\\
		\hline 
		\hskip -1.7 cm I (b) :  Seesaw& \hskip - 4.6cm
		{\scriptsize $  M_D\sim \begin{pmatrix}
			1 & 0 &0 \\  0 & 0 & 1 \\ 0& 1 & 0\end{pmatrix} $}, \quad
		{\scriptsize	$M_N\sim\begin{pmatrix}
			2Y_1 & -Y_3 & -Y_2 \\
			-Y_3 & 2Y_2 & -Y_1 \\
			-Y_2 & -Y_1 & 2Y_3\end{pmatrix}$}\\
		\hline 
		I\hskip -0.03cmI : Weinberg Operator& \hskip -3.5 cm
		{\scriptsize $M_\nu\sim \begin{pmatrix}
			2Y_1 & -Y_3 & -Y_2 \\
			-Y_3 & 2Y_2 & -Y_1 \\
			-Y_2 & -Y_1 & 2Y_3\end{pmatrix}$}\\
		\hline 
		\hskip -0.8 cm I\hskip -0.03cm I\hskip -0.03cmI : Dirac Neutrino &
		{\scriptsize $ M_\nu\sim\begin{pmatrix}
			2g_1Y_1 & (-g_1+g_2)Y_3 & (-g_1-g_2)Y_2 \\
			(-g_1-g_2)Y_3 & 2g_1Y_2 & (-g_1+g_2)Y_1 \\
			(-g_1+g_2)Y_2 & (-g_1-g_2)Y_1 & 2g_1Y_3\end{pmatrix} $}\\ \hline
	\end{tabular}
	\caption{
		The classification of the modular invariant  mass matrices 
		for  neutrino models.}
	\label{tb:neutrino}
\end{table}
For the case where neutrino masses originate from
the Weinberg operator, the superpotential in Eq.(\ref{Weinberg})
is written as:
\begin{align}
\begin{aligned}
w_\nu=&-\frac{v_u^2}{\Lambda}\begin{pmatrix}
2\nu_e\nu_e-\nu_\mu\nu_\tau-\nu_\tau\nu_\mu\\
2\nu_\tau\nu_\tau-\nu_e\nu_\mu-\nu_\mu\nu_\tau\\
2\nu_\mu\nu_\mu-\nu_\tau\nu_e-\nu_e\nu_\tau\end{pmatrix}\otimes
\begin{pmatrix}Y_{1}\\Y_{2}\\Y_{3}\end{pmatrix} \\
=&-\frac{v_u^2}{\Lambda}
\left[(2\nu_e\nu_e-\nu_\mu\nu_\tau-\nu_\tau\nu_\mu)Y_1+
(2\nu_\tau\nu_\tau-\nu_e\nu_\mu-\nu_\mu\nu_e)Y_3
+(2\nu_\mu\nu_\mu-\nu_\tau\nu_e-\nu_e\nu_\tau)Y_2\right].
\end{aligned}
\end{align}
The Majorana neutrino mass matrix is given as follows:
\begin{align}
M_\nu=-\frac{v_u^2}{\Lambda}\begin{pmatrix}
2Y_1 & -Y_3 & -Y_2 \\
-Y_3 & 2Y_2 & -Y_1 \\
-Y_2 & -Y_1 & 2Y_3\end{pmatrix}_{LL}.
\end{align}
This matrix is the same one as in Eq.(\ref{MajoranaR}) 
 apart from the normalization
because both left-handed neutrinos and the right-handed neutrinos
are the triplet of $A_4$.

For the case where the neutrino is the Dirac particle,
we use the mass matrix in Eq.(\ref{MD}).

It is important to  address  the transformation needed to put kinetic terms of matter superfields in the canonical form because kinetic terms are given in  Eq.(\ref{kinetic}).
 The canonical form  is realized by the overall normalization of the lepton mass matrices, which
  shifts our parameters  such as
\begin{eqnarray}
&&\alpha \rightarrow \alpha'= \alpha  (K_{L} K_{e_R}) ^{-1/2}, \qquad
\beta  \rightarrow \beta' = \beta  (K_{L} K_{\mu_R} )^{-1/2}, \qquad
\gamma  \rightarrow \gamma' = \gamma  (K_{L} K_{\tau_R} )^{-1/2}, \nonumber \\
&&g_i \rightarrow  g_i' = g_i (K_{L} K_{\nu_R} )^{-1/2}\ (i=1,2) , \qquad
\Lambda \rightarrow \Lambda' = \Lambda { K_{\nu_R} }^{-1},
\end{eqnarray}
where $K_{\phi}$ denotes a coefficient of the kinetic term of Eq.(\ref{kinetic}).
Hereafter, we rewrite $\alpha$, $\beta$, $\gamma$, $g_i$, and $\Lambda$ 
  for $\alpha'$, $\beta'$, $\gamma'$, $g_i'$, and $\Lambda'$ in our convention. 

 Finally, we summarize the classification of mass matrices
 for neutrino models in Table \ref{tb:neutrino}.
\section{Numerical results}
We discuss numerical results for neutrino models  in  Table \ref{tb:neutrino}.
The lepton mass matrices in the previous section 
 are given by modulus parameter $\tau$.
 By fixing $\tau$, the modular invariance is broken, and then the lepton mass matrices give the mass eigenvalues and flavor mixing numerically.
In order to fix the value of  $\tau$, we use the result of NuFIT 3.2 with the $3\,\sigma$ error-bar \cite{NuFIT}.
We consider both the normal hierarchy (NH) of neutrino masses $m_1<m_2<m_3$ 
and the inverted hierarchy (IH) of neutrino masses $m_3<m_1<m_2$, where
$m_1$, $m_2$, and $m_3$ denote three light neutrino masses.
The sum of neutrino masses are restricted 
by the  cosmological observations~\cite{Giusarma:2016phn,Vagnozzi:2017ovm}.
Planck 2018 results provide us its cosmological upper bound for sum of neutrino masses; 
$120$-$160$~meV \cite{Aghanim:2018eyx} 
at the $95\%$ C.L. depending on the combined data.
We have used the upper bound of $160$~meV as a conservative constraint of our models.
By inputting the data of $\Delta m_{\rm atm}^2 \equiv m_3^2-m_1^2$, $\Delta m_{\rm sol}^2 \equiv  m_2^2-m_1^2$, and 
three mixing angles $\theta_{23}$, $\theta_{12}$, and $\theta_{13}$ with $3\,\sigma$ error-bar   given in Table \ref{DataNufit}, 
we fix the modulus $\tau$ and the other parameters. 
Then we can predict the CP violating Dirac phases $\delta_{CP}$ and
Majorana phases  $\alpha_{31}$,  $\alpha_{21}$, which are defined in Appendix C.

\begin{table}[hbtp]
	\begin{center}
		\begin{tabular}{|c|c|c|}
			\hline 
			\rule[14pt]{0pt}{0pt}
			\  observable \ &  $3\,\sigma$ range for NH  & $3\,\sigma$ range for IH \\
			\hline 
			\rule[14pt]{0pt}{0pt}
			$\Delta m_{\rm atm}^2$& \ \   \ \ $(2.399$ - $2.593) \times 10^{-3}{\rm eV}^2$ \ \ \ \
			&\ \ $(-2.562$ - $-2.369) \times 10^{-3}{\rm eV}^2$ \ \  \\
			\hline 
			\rule[14pt]{0pt}{0pt}
			$\Delta m_{\rm sol }^2$&   $(6.80$ - $8.02)  \times 10^{-5}{\rm eV}^2$
			& $(6.80$ - $8.02)  \times 10^{-5}{\rm eV}^2$ \\
			\hline 
			\rule[14pt]{0pt}{0pt}
			$\sin^2\theta_{23}$&  $0.418$ - $0.613$ & $0.435$ - $0.616$ \\
			\hline 
			\rule[14pt]{0pt}{0pt}
			$\sin^2\theta_{12}$& $0.272$ - $0.346$ & $0.272$ - $0.346$ \\
			\hline 
			\rule[14pt]{0pt}{0pt}
			$\sin^2\theta_{13}$& $0.01981$ - $0.02436$ & $0.02006$ - $0.02452$ \\
			\hline 
		\end{tabular}
		\caption{The $3\,\sigma$ ranges of neutrino oscillation parameters from NuFIT 3.2
			for NH and IH \cite{NuFIT}. }
		\label{DataNufit}
	\end{center}
\end{table}


\subsection{Model I(a): Seesaw }

The coefficients $\alpha/\gamma$ and  $\beta/\gamma$ in the charged lepton mass matrix are given only in terms of $\tau$ after inputting the observed values  $m_e/m_\tau$ and $m_\mu/m_\tau$ as shown in Appendix B.
Then, we have two free parameters, $g_1/g_2$ and the modulus $\tau$ apart from the overall factors in the neutrino sector.
Since these are complex, we set
\begin{equation}
    \tau= {\rm Re}[\tau]+ i\ {\rm Im}[\tau]\ , \qquad\quad \frac{g_2}{g_1}=g\ e^{i \phi_g}\  .
\end{equation}
The fundamental domain of $\tau$ is presented in Ref.\cite{Feruglio:2017spp}.
In practice, we restrict our parametric search in ${\rm Re}[\tau]\in[-1.5,1.5]$ and ${\rm Im}[\tau]>0.6$.
We also  take  $\phi_g\in[-\pi,\pi]$.
These four parameters are fixed by 
the observed $\Delta m_{\rm sol}^2/\Delta m_{\rm atm}^2$ and  three mixing angles $\theta_{23}$, $\theta_{12}$ and $\theta_{13}$.

\begin{figure}[h!]
	\begin{tabular}{ccc}
	\begin{minipage}{0.475\hsize}
		\includegraphics[width=\linewidth]{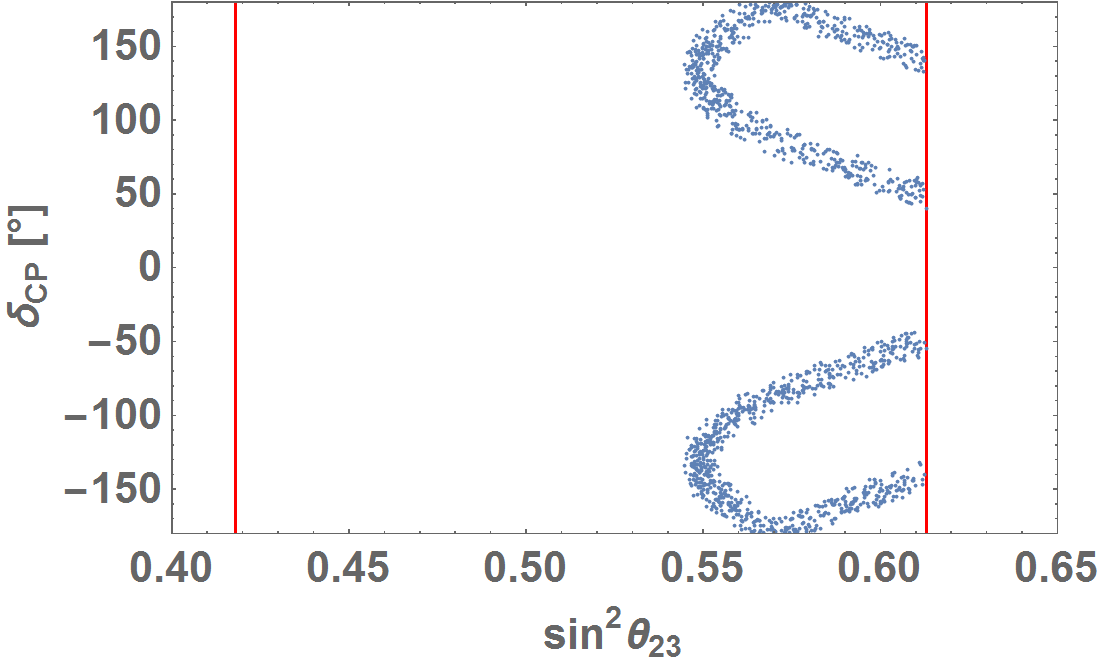}
		\caption{The prediction of $\delta_{CP}$ versus $\sin^2\theta_{23}$ for NH in model I(a).
			The vertical red lines represent the upper and lower bounds of the experimental data with $3 \ \sigma$.}
	\end{minipage}
	\phantom{=}
	\begin{minipage}{0.475\linewidth}
		\includegraphics[width=\linewidth]{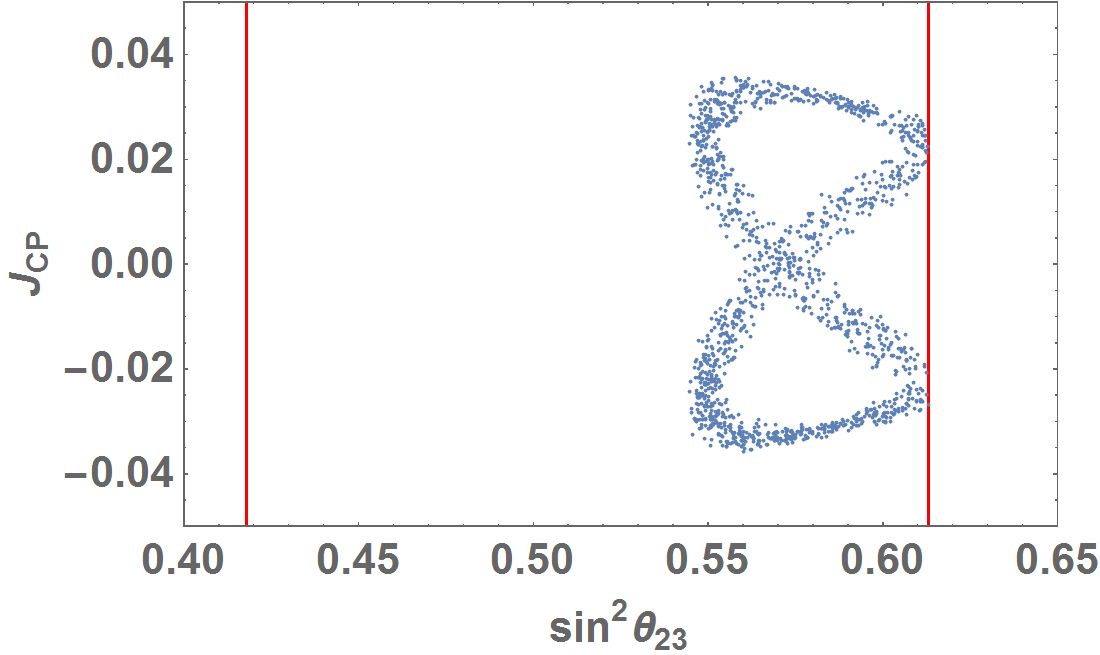}
		\caption{The prediction of $J_{CP}$ versus $\sin^2\theta_{23}$ for NH in model I(a).
		The vertical red lines represent the upper and lower bounds of the experimental data with $3 \ \sigma$.}
	\end{minipage}
	\end{tabular}
\end{figure}
\begin{figure}[h!]
	\begin{tabular}{ccc}
	\begin{minipage}{0.475\hsize}
	\begin{center}
		\includegraphics[width=\linewidth]{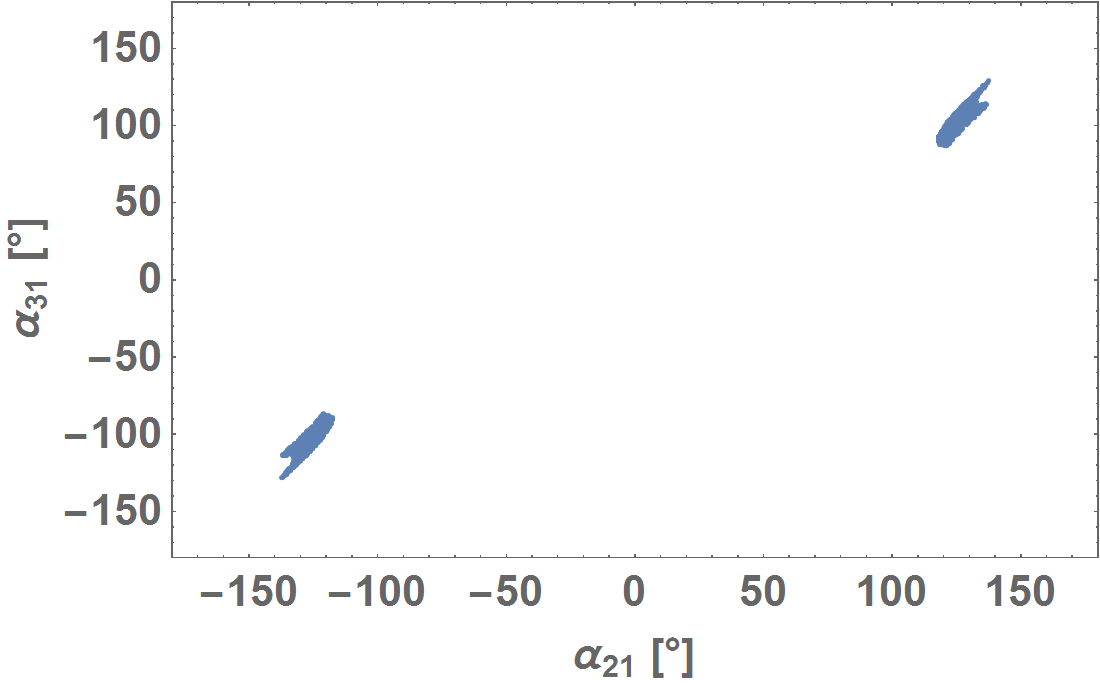}
		\caption{The prediction of Majorana phases $\alpha_{21}$ and $\alpha_{31}$
			for NH in model I(a).}
	\end{center}	
	\end{minipage}
	\phantom{=}
	\begin{minipage}{0.475\hsize}
	\begin{center}
	\vspace{0.5cm}
		\includegraphics[width=\linewidth]{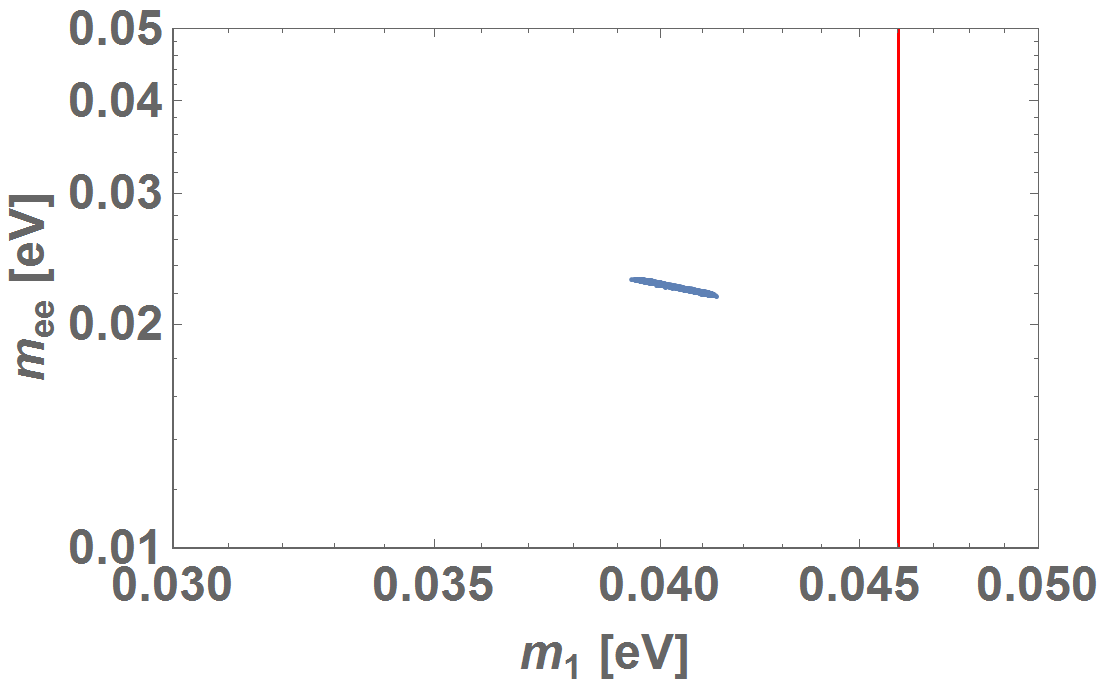}
		\caption{The prediction of $m_{ee}$ versus $m_1$  for NH in model I(a).
			The red vertical line denotes the upper-bound of $m_1$.}
	\end{center}
	\end{minipage}
	\end{tabular}
\end{figure}

\begin{table}[h]
	\centering
	\begin{tabular}{|c|c|c|c|c|c|} \hline 
		\rule[14pt]{0pt}{0pt}
	${\rm Im} [\tau]$	&	${\rm Re} [\tau]$ &$g$ &$\phi_g$& $\alpha/\gamma$ &$\beta/\gamma$ \\ \hline \hline 	\rule[14pt]{0pt}{0pt}	
	$0.66$\,--\,$0.73$&$\pm(0.25\hbox{\,--\,}0.31)$, $\pm(0.46\hbox{\,--\,}0.54)$,&$1.20$\,--\,$1.22$&$\pm (87$\,--\,$88)^\circ$ & $202$\,--\,$203$ &$3286$\,--\,$3306$\\ 
        $1.17$\,--\,$1.32$& $\pm(0.66\hbox{\,--\,}0.75)$, $\pm(1.25\hbox{\,--\,}1.31)$,& &$\pm (92$\,--\,$93)^\circ$	&  &\\
         & $\pm(1.46\hbox{\,--\,}1.50)$& &	&  &\\
        \hline
	\end{tabular}
	\caption{The parameter regions consistent with the experimental data
		of Table \ref{DataNufit} for model I(a).
		Results do not change under the exchange of $\alpha/\gamma$ and $\beta/\gamma$.}
	\label{parameters}
\end{table}


At first, we present the prediction of the Dirac CP violating phase $\delta_{CP}$ versus $\sin^2\theta_{23}$  for NH of neutrino masses in Fig.1.
It is emphasized that $\sin^2\theta_{23}$ is restricted to be larger than $0.54$,
and $\delta_{CP}=\pm (50^{\circ}\mbox{--}180^{\circ})$.
Since the correlation of $\sin^2\theta_{23}$ and $\delta_{CP}$ is characteristic,
this prediction is testable in the future experiments of neutrinos.
On the other hand, predicted $\sin^2\theta_{12}$ and $\sin^2\theta_{13}$ cover
observed full region with $3\,\sigma$ error-bar, and there are no correlations with $\delta_{CP}$.

We also show the predicted Jarlskog invariant $J_{CP}$~\cite{Jarlskog:1985ht},
characterizing the magnitude of CP violation in neutrino oscillations,
versus $\sin^2\theta_{23}$ for NH of neutrino masses in Fig.2.
The magnitude of $J_{CP}$ is predicted to be $0$\,--\,$0.035$ depending on $\theta_{23}$.

We show the prediction of Majorana phases $\alpha_{21}$ and $\alpha_{31}$
in Fig.3. The predicted regions are restricted in $\alpha_{21}= \pm (118^\circ$--$138^\circ $) and  $\alpha_{31}= \pm (86^\circ$--$130^\circ $).
This result is used in the calculation of neutrinoless double beta decay.

Let us show the prediction of  the effective mass $m_{ee}$ which is
the measure of  the neutrinoless double beta decay as seen in Appendix C.
The prediction of $m_{ee}$ is presented versus $m_1 $  in Fig.4.
It is remarkable that $m_{ee}$ is around $22$\,meV while $m_1$ is $40$\,meV.
The red vertical line in Fig.4 denotes the upper bound of $m_1$,
which is derived from the cosmological bound $\sum m_i<160$\,meV.
The obtained value of $m_1$ indicates near degenerate neutrino mass spectrum,
$m_1\simeq m_2\simeq 40$\,meV and $m_3\simeq 60$\,meV.
The prediction of $m_{ee}\simeq 22$\,meV is testable in the future experiments
of the neutrinoless double beta decay.
We predict the rather large sum of neutrino masses as $\sum m_i\simeq 145$\,meV,
which is required by consistency with the observed value of $\sin^2\theta_{13}$.

The parameters of our model  are determined by the input data of Table \ref{DataNufit}.
Numerical values are listed in Table \ref{parameters}. 
 


We have also scanned the parameter space for the case of IH of neutrino masses.
We have found parameter sets which fit the data of  $\Delta m_{\rm sol}^2$, $\Delta m_{\rm atm}^2$
and  three mixing angles $\sin^2\theta_{23}$, $\sin^2\theta_{12}$, and $\sin^2\theta_{13}$.
However, the predicted $\sum m_i$ is around $190$--$200$\,meV. 
Therefore, we also omit to show numerical results.

\subsection{Model I(b): Seesaw}

There is another assignment of the modular weight for the left-handed lepton and the right-handed charged leptons as presented in parentheses of Table \ref{tb:fields}
\cite{Criado:2018thu}.
Then, the Dirac neutrino mass matrix is given by the constant parameter
as seen in Eq.(\ref{Dirac-constant}).
We have scanned the parameter space for both NH and  IH of neutrino masses.
The parameters to reproduce the observed $\Delta m_{\rm sol}^2$ and $\Delta m_{\rm atm}^2$
cannot give the large mixing angle of $\theta_{23}$.
The predicted value $\sin^2\theta_{23}\simeq 0.18$ for NH.
We also obtain $\sin^2\theta_{12}\simeq 0.8$ and  $\sin^2\theta_{13}\simeq 0.15$.
On the other hand, the predicted value $\sin^2\theta_{23}\simeq 0$, $\sin^2\theta_{12}\simeq 0.5$, and $\sin^2\theta_{13}\simeq 0$ for IH.
In conclusion, the model I(b) is inconsistent with the experimental data of Table \ref{DataNufit}.
     
It may be useful to add the discussion on the model by Criado and Feruglio
\cite{Criado:2018thu}, where the charged lepton mass matrix is
different from ours in Eq.(\ref{eq:CL}), but given by a flavon while the neutrino mass matrix  is just same one in model I(b).
We have reproduced the numerical results
of Ref. \cite{Criado:2018thu}, in which the three mixing angles and masses
are consistent with the experimental data 
and the cosmological bound, respectively, for  NH of neutrino masses.
The predicted CP violating phase is $\delta_{CP}\simeq \pm 100^\circ$.

\subsection{Model I\hspace{-.1em}I: Weinberg Operator}

In this case, the modulus $\tau$ is the  only parameter in the neutrino mass matrix apart from the overall factors.
We can find the parameter space to be consistent with the observed  $\sin^2\theta_{12}$
as well as $\Delta m_{\rm sol}^2$ and $\Delta m_{\rm atm}^2$ for both NH and IH.
However, the predicted  $\sin^2\theta_{23}$ is around $0.8$ and $\sin^2\theta_{13}$ is very large as  $0.45$ for NH.
On the other hand, for IH, the predicted  $\sin^2\theta_{23}$ is rather small as $0.35$ and
$\sin^2\theta_{13}$ is around $0.04$, which is larger than $1.6$ times of the observed value.
Thus, the neutrino mass matrix by the Weinberg operator do not lead to  the realistic flavor mixing.
  
\subsection{Model I\hspace{-.1em}I\hspace{-.1em}I:  Dirac Neutrino}

There is still a  possibility of  the neutrino being the Dirac particle.
Then, the neutrino mass matrix is different from the Majorana one as shown in Table \ref{tb:neutrino}
although parameters are $\tau$ and $g$ likewise in the case of the seesaw model I(a).

We have found the parameter space to be consistent with both observed $\sin^2\theta_{23}$ and $\sin^2\theta_{12}$
as well as $\Delta m_{\rm sol}^2$ and $\Delta m_{\rm atm}^2$ for NH.
However, the predicted $\sin^2\theta_{13}$ is much smaller than the observed value of ${\cal O}(10^{-3})$.
  
\begin{figure}[h!]
	\begin{tabular}{ccc}
	\begin{minipage}{0.475\hsize}
		\includegraphics[width=1.03\linewidth]{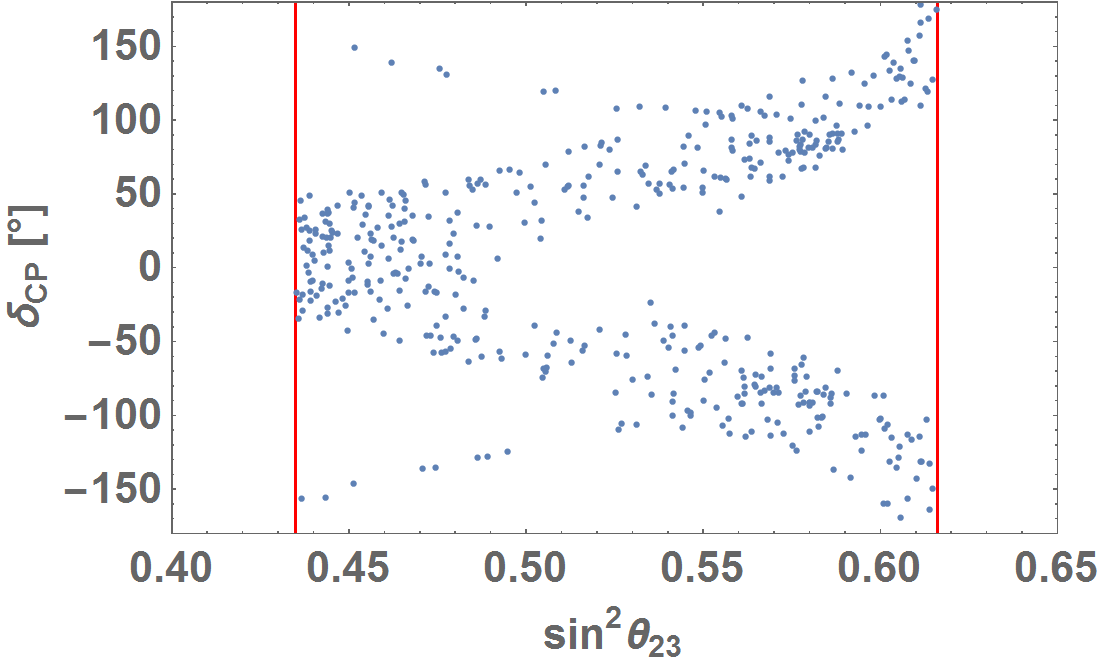}
		\caption{The prediction of $\delta_{CP}$ versus $\sin^2\theta_{23}$  for IH in model I\hskip -0.03cm I\hskip -0.03cm I.
			The vertical red lines represent the upper and lower bounds of the experimental data with $3 \ \sigma$.}	
	\end{minipage}
	\phantom{=}
	\begin{minipage}{0.475\hsize}
		\includegraphics[width=\linewidth]{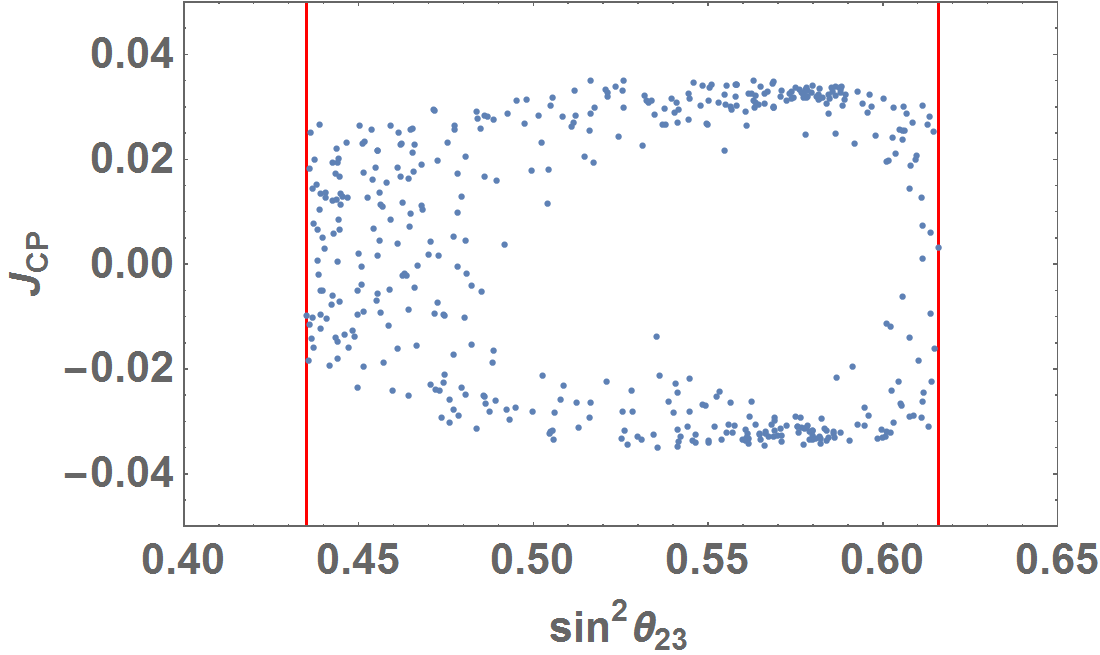}
		\caption{The prediction of $J_{CP}$ versus $\sin^2\theta_{23}$ for IH in model I\hskip -0.03cm I\hskip -0.03cm I.
		The vertical red lines represent the upper and lower bounds of the experimental data with $3 \ \sigma$.}
	\end{minipage}
	\end{tabular}
\end{figure}
\begin{table}[h]
	\centering
	\begin{tabular}{|c|c|c|c|c|c|} \hline 
		\rule[14pt]{0pt}{0pt}
		${\rm Im} [\tau]$	&	${\rm Re} [\tau]$ &$g$ &$\phi_g$& $\alpha/\gamma$ &$\beta/\gamma$ \\ \hline \hline 
		\rule[14pt]{0pt}{0pt}	
		$0.90$\,--\,$1.12$ & $\pm(0.01$\,--\,$0.07)$&$1.43$\,--\,$2.12$&$\pm (76$\,--\,$104)^\circ$
		&$59$\,--\,$88$&$857$\,--\,$1302$\\ 
		&$\pm(0.94$\,--\,$1.10)$& & 	&  &\\ 	
		\hline
	\end{tabular}
	\caption{The parameter regions consistent with the experimental data
		of Table \ref{DataNufit} for model I\hskip -0.03cm I\hskip -0.03cm I.	Results do not change under the exchange of $\alpha/\gamma$ and $\beta/\gamma$. }
	\label{parameters2}
\end{table}

On the other hand,   the  $\sin^2\theta_{13}$ is
completely consistent with  the observed value for IH of neutrino masses.
We present the prediction of the Dirac CP violating phase $\delta_{CP}$ versus $\sin^2\theta_{23}$  for  IH in Fig.5.
The predicted  $\delta_{CP}$ is still allowed in $[-\pi,\pi]$
depending on the magnitude of  $\sin^2\theta_{23}$. 
Since there are no correlations
among $\sin^2\theta_{12}$,  $\sin^2\theta_{13}$, and $\delta_{CP}$,
we omit figures of  $\sin^2\theta_{12}$ and  $\sin^2\theta_{13}$.
    
We also show the predicted Jarlskog invariant $J_{CP}$
versus $\sin^2\theta_{23}$ for IH of neutrino masses in Fig.6.
The magnitude of $J_{CP}$ is predicted to be $0$\,--\,$0.035$.

The  $\sum m_i$ is required in $102\mbox{\,--\,}150$\,meV to be consistent with 
the observed value of   $\sin^2\theta_{13}$.

We summarize numerical values of  parameters  in Table 5.
In the Dirac neutrino model, the neutrinoless double beta decay is forbidden.


\section{Summary}

We study the phenomenological implications of 
the modular symmetry $\Gamma(3) \simeq A_4$
facing recent experimental data of neutrino oscillations.
The  mass matrices of neutrinos and charged leptons
are essentially given  by fixing the expectation value of the  modulus $\tau$,
which is the only source of modular invariance breaking.
We introduce no flavons in contrast with conventional flavor models with the $A_4$ symmetry.
We classify the neutrino models along with type I seesaw (model I(a) and I(b)),
Weinberg operator (model I\hspace{-.1em}I), and Dirac neutrino (model I\hspace{-.1em}I\hspace{-.1em}I).
For the charged lepton mass matrix, three right-handed charged leptons $e_R$, $\mu _R$, and $\tau _R$ 
are assigned to three different singlets $1$, $1''$, and  $1'$ of $A_4$, respectively.

For NH of neutrino masses, we have found that the  seesaw model I(a) is available
facing recent experimental data of NuFIT 3.2 \cite{NuFIT}
and the cosmological bound of the sum of neutrino masses \cite{Aghanim:2018eyx}.
The predicted  $\sin^2\theta_{23}$ is restricted to be larger than $0.54$
and $\delta_{CP}=\pm (50^{\circ}\mbox{\,--\,}180^{\circ})$. 
The sharp correlation between $\sin^2\theta_{23}$ and $\delta_{CP}$ is testable in the future experiments of the neutrino oscillations.
It is remarkable that $m_{ee}$ is around $22$\,meV while the sum of neutrino masses is $145$\,meV.

For IH of neutrino masses, the Dirac neutrino model I\hskip -0.03cm I\hskip -0.03cm I is completely consistent with
the experimental data of NuFIT 3.2 and the cosmological bound of the sum of neutrino masses.
The predicted  $\delta_{CP}$ is still allowed in $[-\pi,\pi]$ depending on the magnitude of $\sin^2\theta_{23}$. 
The $\sum m_i= 102$\,--\,$150$\,meV is required by consistency with the observed value of $\sin^2\theta_{13}$.

The seesaw model I(b) and the Weinberg operator model I\hskip -0.03cm I cannot reproduce the observed mixing angles
after inputting the data of $\Delta m_{\rm sol}^2$ and $\Delta m_{\rm atm}^2$ for both NH and IH.


It is helpful to comment on the effects of the supersymmetry (SUSY) breaking 
and the radiative corrections because we have discussed our model in the  limit of exact SUSY. 
The SUSY breaking effect can be neglected if the separation between the SUSY breaking scale and the SUSY breaking mediator scale is sufficiently large \cite{Criado:2018thu}.
In our numerical results, the corrections by the renormalization are 
very small as far as we take the relatively small value of $\tan\beta$.

We have focused on the models with Yukawa couplings and masses, which correspond to modular forms of lower weights.
Such models have strong constraints.
Similarly, we can construct models with modular forms of higher weights.
Such models would have a variety in model building.

\vspace{0.5cm}
\noindent

{\bf Acknowledgement}  

We would like to thank F. Feruglio and J.T. Penedo for useful discussions. 
This work is supported by  MEXT KAKENHI Grant Number JP17H05395 (TK), and 
JSPS Grants-in-Aid for Scientific Research 18J10908 (NO)
16J05332 (YS) and 15K05045, 16H00862 (MT), and 18J11233 (THT).


\appendix

\section*{Appendix}

\section{Multiplication rule of $A_4$ group}
\label{sec:multiplication-rule}
We use the multiplication rule of the $A_4$ triplet as follows:
\begin{align}
\begin{pmatrix}
a_1\\
a_2\\
a_3
\end{pmatrix}_{\bf 3}
\otimes 
\begin{pmatrix}
b_1\\
b_2\\
b_3
\end{pmatrix}_{\bf 3}
&=\left (a_1b_1+a_2b_3+a_3b_2\right )_{\bf 1} 
\oplus \left (a_3b_3+a_1b_2+a_2b_1\right )_{{\bf 1}'} \nonumber \\
& \oplus \left (a_2b_2+a_1b_3+a_3b_1\right )_{{\bf 1}''} \nonumber \\
&\oplus \frac13
\begin{pmatrix}
2a_1b_1-a_2b_3-a_3b_2 \\
2a_3b_3-a_1b_2-a_2b_1 \\
2a_2b_2-a_1b_3-a_3b_1
\end{pmatrix}_{{\bf 3}}
\oplus \frac12
\begin{pmatrix}
a_2b_3-a_3b_2 \\
a_1b_2-a_2b_1 \\
a_3b_1-a_1b_3
\end{pmatrix}_{{\bf 3}\  } \ , \nonumber \\
\nonumber \\
{\bf 1} \otimes {\bf 1} = {\bf 1} \ , \qquad &
{\bf 1'} \otimes {\bf 1'} = {\bf 1''} \ , \qquad
{\bf 1''} \otimes {\bf 1''} = {\bf 1'} \ , \qquad
{\bf 1'} \otimes {\bf 1''} = {\bf 1} \  .
\end{align}
More details are shown in the review~\cite{Ishimori:2010au,Ishimori:2012zz}.


\section{Determination of $\alpha/\gamma$ and $\beta/\gamma$}
The coefficients $\alpha$, $\beta$, and $\gamma$ in Eq.(\ref{eq:CL})
are taken to be real positive without loss of generality.
We show these parameters are described in terms of the modular parameter $\tau$ and the charged lepton masses.
We rewrite the mass matrix of Eq.(\ref{eq:CL}) as  
\begin{align}
M_E^{(1)}&=\gamma Y_3{\rm diag}[\hat{\alpha}, \hat{\beta}, 1]
\begin{pmatrix}
\hat{Y_1} & \hat{Y_2} & 1 \\
1 &  \hat{Y_1} &\hat{Y_2} \\
\hat{Y_2} & 1 &  \hat{Y_1}
\end{pmatrix}_{RL},
\end{align}
where $\hat{\alpha}\equiv\alpha/\gamma$, $\hat{\beta}\equiv\beta/\gamma$, $\hat{Y_1}\equiv Y_1/Y_3$, and $\hat{Y_2}\equiv Y_2/Y_3$.
We use the relation $Y_2^2+2Y_1Y_3=0$ to eliminate $Y_1$ in the equation.
Then, we obtain the following three equations:
\begin{align}
{\rm Tr}[M_E^{(1)\dag} M_E^{(1)}]=\sum_{i=e}^\tau m_i^2&=\frac{|\gamma Y_3|^2}4(1+\hat\alpha^2+\hat\beta^2)C_1~,\label{eq:sum} \\
{\rm Det}[M_E^{(1)\dag} M_E^{(1)}]=\prod_{i=e}^\tau m_i^2&=\frac{|\gamma Y_3|^6}{64}\hat\alpha^2\hat\beta^2C_2~, \label{eq:prod}\\
\frac{{\rm Tr}[M_E^{(1)\dag} M_E^{(1)}]^2-{\rm Tr}[(M_E^{(1)\dag} M_E^{(1)})^2]}2
=\chi&=\frac{|\gamma Y_3|^4}{16}(\hat\alpha^2+\hat\alpha^2\hat\beta^2+\hat\beta^2)C_3~, \label{eq:chi}
\end{align}
where $\chi\equiv m_e^2m_\mu^2+m_\mu^2m_\tau^2+m_\tau^2m_e^2$.
The coefficients $C_1$, $C_2$, and $C_3$ depend only on $\hat{Y_2}\equiv Ye^{i\phi_Y}$, where
$Y$ is real positive and $\phi_Y$ is a phase parameter,
\begin{align}
\begin{aligned}
C_1&=(2+Y^2)^2,\\
C_2&=64+400Y^6+Y^{12}-40Y^3(Y^6-8)\cos(3\phi_Y)-16Y^6\cos(6\phi_Y)~,\\
C_3&=16+16Y^2+36Y^4+4Y^6+Y^8-8Y^3(Y^2-2)\cos(3\phi_Y)~.
\end{aligned}\label{eq:Cs}
\end{align}
These values are determined if the value of modulus $\tau$ is fixed.
Then, we obtain the general equations
which describe $\hat\alpha$ and $\hat\beta$ as functions of charged lepton masses and $\tau$:
\begin{align}
\begin{aligned}
\frac{(1+s)(s+t)}t&=\frac{(\sum m_i^2/C_1)(\chi/C_3)}{\prod m_i^2/C_2}~,\quad\qquad
\frac{(1+s)^2}{s+t}&=\frac{(\sum m_i^2/C_1)^2}{\chi/C_3}~,
\end{aligned}
\end{align}
where we redefine the parameters $\hat\alpha^2+\hat\beta^2=s$ and $\hat\alpha^2\hat\beta^2=t$.
They are related as follows,
\begin{align}
\hat\alpha^2=\frac{s\pm\sqrt{s^2-4t}}2~,\quad\quad
\hat\beta^2=\frac{s\mp\sqrt{s^2-4t}}2~.
\label{alphabeta}
\end{align}

\section{Lepton mixing  and neutrinoless double beta decay}
Supposing neutrinos to be Majorana particles, 
the PMNS matrix $U_{\text{PMNS}}$~\cite{Maki:1962mu,Pontecorvo:1967fh} 
is parametrized in terms of the three mixing angles $\theta _{ij}$ $(i,j=1,2,3;~i<j)$,
one CP violating Dirac phase $\delta _\text{CP}$, and two Majorana phases 
$\alpha_{21}$, $\alpha_{31}$  as follows:
\begin{align}
U_\text{PMNS} =
\begin{pmatrix}
c_{12} c_{13} & s_{12} c_{13} & s_{13}e^{-i\delta _\text{CP}} \\
-s_{12} c_{23} - c_{12} s_{23} s_{13}e^{i\delta _\text{CP}} &
c_{12} c_{23} - s_{12} s_{23} s_{13}e^{i\delta _\text{CP}} & s_{23} c_{13} \\
s_{12} s_{23} - c_{12} c_{23} s_{13}e^{i\delta _\text{CP}} &
-c_{12} s_{23} - s_{12} c_{23} s_{13}e^{i\delta _\text{CP}} & c_{23} c_{13}
\end{pmatrix}
\begin{pmatrix}
1&0 &0 \\
0 & e^{i\frac{\alpha_{21}}{2}} & 0 \\
0 & 0 & e^{i\frac{\alpha_{31}}{2}}
\end{pmatrix},
\label{UPMNS}
\end{align}
where $c_{ij}$ and $s_{ij}$ denote $\cos\theta_{ij}$ and $\sin\theta_{ij}$, respectively.

The rephasing invariant CP violating measure, the Jarlskog invariant~\cite{Jarlskog:1985ht},
is defined by the PMNS matrix elements $U_{\alpha i}$. 
It is written in terms of the mixing angles and the CP violating Dirac phase as:
\begin{equation}
J_{CP}=\text{Im}\left [U_{e1}U_{\mu 2}U_{e2}^\ast U_{\mu 1}^\ast \right ]
=s_{23}c_{23}s_{12}c_{12}s_{13}c_{13}^2\sin \delta _\text{CP}~ .
\label{Jcp}
\end{equation}
There are also other invariants $I_1$ and $I_2$ associated with Majorana phases
\cite{Bilenky:2001rz,Nieves:1987pp,AguilarSaavedra:2000vr,Girardi:2016zwz},
\begin{equation}
I_1=\text{Im}\left [U_{e1}^\ast U_{e2} \right ]
=c_{12}c_{12}c_{13}^2\sin \left (\frac{\alpha_{21}}{2}\right )~, \quad
I_2=\text{Im}\left [U_{e1}^\ast U_{e3} \right ]
=c_{12}s_{13}c_{13}\sin \left (\frac{\alpha_{31}}{2}-\delta_\text{CP}\right )~.
\label{Jcp}
\end{equation}
We calculate $\delta_\text{CP}$, $\alpha_{21}$, and $\alpha_{31}$ with these relations.

In terms of these parametrization, the effective mass for the $0\nu\beta\beta$ decay is given as follows:
\begin{align}
m_{ee}=\left|m_1 c_{12}^2 c_{13}^2+m_2s_{12}^2 c_{13}^2 e^{i\alpha_{21}}+m_3 s_{13}^2 e^{i(\alpha_{31}-2\delta_{CP})}\right|  \ .
\end{align}


\end{document}